*Development of Hardware for Programming of Spatial Three-Dimensional Magnetic Field Distributions in Magnetic Resonance Imaging and High-Resolution Nuclear Magnetic Resonance (a proposal)*

Contents

Introduction

Spatial distribution of the static magnetic field within sample volume was mainly *terra incognita* since the first experiments of Nuclear Magnetic Resonance (NMR) [1-3] until shimming procedure was introduced for some global control of the field distribution. In particular, the use of the shimming coils [4-5] had allowed improvement of the static magnetic field homogeneity from 0.1 to 0.001 ppm by application of globally acting field shapes produced by conducting wires of specific geometry. This was crucial for introduction of high-resolution applications and allowed study of two-dimensional spin density distribution within a sample [6], what led to invention of Magnetic Resonance Imaging (MRI). Also, pulsed field gradients were introduced in NMR [7-8] and application of three-dimensional spatial pulsed field gradients [9] became inciting in expansion of multidimensional NMR techniques to MRI. Active shielding [10] had improved performance of gradient coils in pulsed mode and was followed with introduction of the pulsed field gradient modules. Later progress was made by commencement of novel coil architectures such as active phased arrays [11] and other innovative designs [12-13] which, in particular, offer enhanced control over local field distributions.

However, the limitations of the methods for control of spatial magnetic field distributions in existence are that they are predominantly globally acting and normally have scarce repertory of the available field shapes which is moreover fixed yet at the stage of coils fabrication.

Motivation

These limitations restrict development of shimming of high-order field inhomogeneities [14, 15], creation of local spoilers and production of other complex arbitrary magnetic field shapes. A variety of field shapes is especially useful for optimization of spatial response function in MRI when it's supposed to be used with multiformity of sample shapes without loss of sensitivity and for reduction of artefacts [16]. The magnetic field shape diversity can be particularly useful for simultaneous localization from different voxels within sample volume [17-20]. Availability of the high-order field shapes would contribute into engineering of nonlinear local Larmor precession and nonlinear phase- and frequency-encoding [15] to be exploited in pulse sequence design.

Introduction of the novel spatially various magnetic field shapes will also facilitate the transfer from traditional 'repetitive' to simultaneous NMR and MRI methods [12, 21] where it's relevant. This is achieved at expense of some sensitivity and resolution losses what is acceptable in a range of experiments [21] and promises to reduce experimental time significantly. Such reduction would be crucial for further development of normally time-consuming experiments such as multi-dimensional NMR experiments in structural biology, for example, where many hours and even days are required to complete an experiment. This is also relevant because cost of high-resolution NMR machine time is usually high (in European Union it's normally on scale of about 4000 euros per month [22]).

Hence, the development of the hardware for generation of arbitrary and programmable magnetic field shapes (static and, in principle, high-frequency as well) would be significant for further development of NMR/MRI techniques and sets the aims of this project.

Realization

An approach accepted in this project for development of the magnetic field spatial distribution programming module is described in this section.

Programming of the physical properties assumes their digital control [23, 24] and can be realized using concept of metamaterials [25] interpreted here as embedded systems [26], where hardware is built as discrete periodic structure whose properties of interest can be programmed at low level according to user's instructions. The hardware design in this project will employ the concepts and techniques adopted from the field of *reconfigurable antennas* [27].

In the natural environment, where mainly the Earth's magnetic field NMR/MRI is present, such hardware can be developed using either fixed conducting elements interconnected by switches (on basis of PIN diodes [28] or MEMS [29]), or dynamically programmed conducting paths on basis of silicon structures [30], either as a combination of the both. Although these techniques can be directly exploited for ultra-low field NMR/MRI [31], the most of modern high-resolution NMR spectrometers and MRI devices use strong static magnetic fields what excludes the use of electromagnetic activation for configuration of the conducting elements geometry and, thus, put limits on their use in the strong field environment. Meanwhile, the use of other mechanisms for activation of MEMS switches - electrostrictive, for example, was proved in MRI environment [32] and can be implemented in this project. In general, MEMS are capable to operate in the frequency range from DC to GHz and, therefore, can be used in the hardware aimed on the static and/or high frequency magnetic field shape programming when it is complemented with the relevant circuitry.

At the first stage of the project, custom software for calculation of the conducting paths geometry required to produce specific static field shape within sample volume will be implemented using Turner-Carlson [33-36] and/or IBEM [12, 37-40] approaches to gradient and shim coil design. This software will take the fixed positions of current carrying elements regularly distributed over cylindrical surface as constraint and calculate the interconnections between these elements to form the conducting paths required. These designs will be verified by using Ansoft Maxwell and Ansoft HFSS software packages [41] for electromagnetic modeling. The next step will be a design of hardware (MEMS assembly) using CoventorMEMS+ [42] software package followed by its fabrication in the MEMS Foundries [43]. In parallel to these stages, the custom firmware will be designed using Xilinx ISE and EDK [44, 45] software packages. An interface between Xilinx FPGA and user's computer will be implemented using custom-made AVR microcontroller-based device [46, 47]. These microcontroller and FPGA-based modules will provide digital control of the MEMS assembly.

An aim of the final stage of the project will be testing of the device which can be done using low budget means, for example - spectrometer iSpin [48] equipped with a magnet Model4S from SpinCore [49], and added by a probe specially designed as described [50]. This testing can be added (when required) with improvements and modifications.

Risks *versus* benefits expected

This project is conceptual in nature and, hence, main uncertainties are associated with that how exactly proposed approach meets aims of the project. Although the developments described in the previous chapter are in favor of the general approach there are technical challenges associated with the project which can potentially turn into limiting factors for applicability of the device proposed. In particular, it will be important to find out experimentally the precision with which spatial resolution of the local field distributions can be programmed for static field shapes. A need for active shielding for pulsed mode operation has to be taken into account in the design process. The bounds of the available power supplies may also be a limiting factor to some extent. Although it's believed that most of these challenges can be resolved in practice some bounds are likely to stay limiting factors to some extent anyway by the end of the project and possibility of their reduction will be of special concern throughout the project.

The main benefit expected from this project is that flexibility and programmability of the static magnetic field shapes can be added to NMR and MRI most thoroughly for the first time. Another benefit is associated with possibility of further expansion of the approach to spatial high-frequency magnetic field shapes generation which would be natural as MEMS, which will be used in the project for reconfiguration of the coils geometries, can operate up to microwaves. Another benefit emerges from variety of the MEMS sizes available, what (with some limitations) makes the design of the device proposed scalable for whole range of sample sizes under study at present - from microlitre to whole-body MRI applications.

Information about author

Name: Vladimir Korostelev.
E-mail: microgradientrussia@gmail.com

In a period from 2001 till 2004 I was working under supervision of Professor Gareth Morris in University of Manchester, UK on a Ph.D. project aimed on improvements of three dimensional automated shimming techniques in high resolution NMR. This project was carried out in collaboration with Varian Inc. The routines developed during this project were included in standard Vnmr software package for 3D automated shimming (starting from 6.1G version).